# Quasi-unit cell description of two-dimensional octagonal quasilattice

Longguang Liao, Xiujun Fu[*], Zhilin Hou

*Department of Physics, South China University of Technology, Guangzhou 510640, China*

**Abstract**

We present a cluster covering scheme to construct the two-dimensional octagonal quasilattice. A quasi-unit cell is successfully found which is a two-color cluster similar to the Gummelt's two-color decagon in five-fold quasilattice. The quasi-unit cells overlap each other following certain covering rules and thus lead to a perfect octagonal quasilattice.

---

[*] Corresponding author.
 Tel.: +86-20-8711-3934; fax: +86-20-8711-4566
 E-mail address: phxjfu@scut.edu.cn (X. Fu)

1. **Introduction**

   Research on quasicrystals started two decades ago soon after the first experimental observation that diffraction pattern with five-fold symmetry appeared in a metallic phase [1]. Quasicrystals are solids which possess quasiperiodic translational order and rotational symmetries that are forbidden in traditional periodic structures [2]. An ideal quasicrystal is constructed by the infinite repetition in space of two or more distinct atomic unit cells. Therefore, quasicrystal models are often described by tiling of two or more building blocks. In two dimensions, the most famous model is the Penrose tiling [3] which consists of two types of rhombus, one with acute angle $2\pi/5$ and another with acute angle $\pi/5$. When the rhombi are packed according to some specific matching rules, a two-dimensional Penrose quasilattice is obtained.

   The theoretical explanation for the formation and stabilization of quasicrystals is still a problem which is poorly understood. Compared with the single unit cell picture for periodic crystals, a quasiperiodic tiling needs at least two distinct unit cells to fill a space, which seems difficult to realize in physical experiment. Therefore, it is necessary to develop other models for quasicrystals. Cluster covering is a new approach to form quasiperiodic structures. Jeong and Steinhardt [4] have shown that a Penrose tiling can be generated by maximizing the density of a particular tile cluster. The cluster is composed of several fat and thin rhombi which represents a low-energy, microscopic cluster of atoms. And overlapping clusters will form ordered structures. On the other hand, Gummelt [5] has proposed a decagonal covering scheme from a purely geometric point of view and has shown that a quasiperiodic structure equivalent to Penrose tiling can be obtained through the covering of a single type of decagon. The Gummelt's covering method needs only one type of prototile which is a decorated decagon. When the decagons overlap each other according to certain rules, a quasiperiodic pattern is formed. In this sense, the Gummelt's two-color decagon can be considered as a quasi-unit cell (QUC), which is similar to the concept of unit cell for periodic crystals.

   The cluster covering theory has attracted much attention in recent years, and most of the publications focused on the decagonal covering quasicrystals [6-14]. A question naturally arises: Can the cluster covering scheme be used to describe other quasilattices such as the octagonal tiling [15,16]? Gähler and his coworkers [17-20] have investigated the octagonal Ammann-Beenker tiling by the cluster approach. It has been shown that such a quasiperiodic structure can be constructed in two ways. The first one [17] is based on the cluster maximization principle where an octagon and a ship are chosen as shown in Fig.1(a). By maximizing the density of these two clusters with suitable relative weights, they obtained the octagonal tiling with eight-fold symmetry. The second way [18-20] to get the octagonal quasilattice is also based on the cluster maximization principle, but it only uses a single cluster which consists of two squares and four rhombi and each of their edges is assigned an arrow [Fig.1(b)]. A tiling completely covered by the arrowed clusters has to satisfy the alternation condition which forces to the Ammann-Beenker tiling. The arrowed octagon can also be replaced by an undecorated cluster of Fig.1(c) but in this case the cluster contains 18 squares and 28 rhombi, which seems more complicated.

In this work we will explore the QUC description for the octagonal Ammann-Beenker tiling following the Gummelt's covering idea of two-color decagon. We begin with a dart-triangle octagonal tiling to find out the fundamental cluster that can cover the whole pattern. By a careful analysis, the QUC is successfully constructed. The covering rules between two QUCs are also presented.

## 2. The quasi-unit cell

The octagonal Ammann-Beenker lattice is a two-dimensional quasicrystal model with eight-fold rotational symmetry [15,16]. It is composed of two tiles, a square and a $45^0$ rhombus. Another version of the octagonal quasilattice is the dart-triangle tiling [17] which is shown in Fig.2. It is this pattern which we base on to derive the QUC.

The dart and triangle tiles touch edge to edge in the quasilattice. What we are looking for is a fundamental cluster consisting of several darts and triangles that can form the same pattern through covering. From Fig.2 we first note that there are many overlapped hexadecagons, each of which consists of 36 triangles and 10 darts. This hexadecagon should be the main part of the QUC and we call it *A* cluster. However, when eight *A* clusters overlap in the way shown in Fig.2(a) a vacancy appears at the center. Therefore, some other things must be added to the hexadecagon.

It is observed that there are always eight octagons surrounding an *A* cluster. Each octagon is made of eight isosceles triangles. If the eight octagons are considered as part of the QUC, then the above vacancy is filled. We thus get the *B* cluster as shown in Fig.2(b).

There still exist other vacancies when two *B* clusters overlap in some ways. As shown in Fig.2(c), two triangle vacancies appear when clusters *B1* and *B2* cover each other. And one vacancy shows up which consists of two triangles and one dart when clusters *B2* and *B3* cover each other. These vacancies do not belong to any other *B* clusters. After attaching one dart and one triangle between any two outer octagons of the *B* cluster, we get the *C* cluster shown in Fig.2(d). This is the whole body of our QUC.

The next work is to color the *C* cluster. Now there are eleven octagons consisting of eight triangles in a *C* cluster, three inside the hexadecagon and eight outside the hexadecagon. When two *C* clusters cover each other, they share two or more octagons. So we color the decagons with black.

Since there is a correspondence between the Ammann-Beenker tiling and the covering pattern, the asymmetry in the former should be reflected in the latter. By connecting the centers of the eleven octagons, we find a Gähler's octagon is formed [Fig.3(a)]. If arrows are added to the edges of squares and rhombi according to the Gähler's octagonal cluster, it is found that the direction of each arrow coincides with the direction of the dart lying on the arrowed edges.

If the eighteen darts are labeled 1 to 18 as in Fig.3(a), then darts 1 to 16 lie on the edges of Gähler's cluster but darts 17 and 18 lie inside the two squares. Taking all the covering cases of *C* clusters into account, it is found that the darts 1 to 16 can overlap each other, but darts 17 and 18 cannot overlap with any one of darts 1 to16. Therefore, we color darts 1-16 with black, same as the octagons. The rest of the *C* cluster keeps white. This finally results in the QUC as shown in Fig.3(b).

The rightness of the QUC can be confirmed in the ideal octagonal covering pattern. We generate a large patch of Ammann-Beenker tiling which contains more than 30,000 triangles and darts and color them according to the above scheme. We have double checked the coverings of the QUC and no mis-matching of color is found. A patch of the covering pattern of QUC is shown in Fig.4, where the eight-fold symmetry can be clearly seen.

### 3. The geometric parameters of QUC and covering cases

As we can see from Fig.3 that there are eleven black parts in the QUC, each of which originates from an octagon in *C* cluster. Now we locate the positions of the centers of the octagons. Let *O* be the center of the C cluster and *A* to *K* be respectively the centers of the eleven octagons [Fig.3(a)]. If we set the spacing between any two centers of adjacent outer octagons to be unit, i.e. $AB = BC = ... = HA = 1$, then some specific geometric parameters are obtained as follows.

$$OA = OB = ... = OH = \sqrt{2}\cos\frac{\pi}{8} \simeq 1.307,$$

$$OI = OJ = OK = \sqrt{2}\sin\frac{\pi}{8} \simeq 0.541,$$

$$JB = 2\sin\frac{\pi}{8} \simeq 0.765.$$

The size of the octagon is characterized by

$$BL = \tan\frac{\pi}{8} = \sqrt{2} - 1 \simeq 0.414.$$

The *A* cluster is not a regular hexadecagon which has two edge lengths corresponding to the base and the leg of an isosceles triangle with the length ratio $2\sin(\pi/8):1 \simeq 0.765$.

Now we present the covering rule of the QUC: two QUCs cover each other only if they share the same color and the overlapped region is greater than or equal to the area of two black octagons of Fig.3(b). Under this constraint, there are 43 possible covering cases between two QUCs. However, not all of them can survive in a perfect octagonal quasilattice due to the requirement of the long-range quasiperiodic translational order. The total number of allowed coverings is 29 and all the covering cases are shown in Fig.5.

To describe the 29 types of coverings in an accurate way, we need to specify the relative position and orientation of two QUCs. First, we use three parameters $(x_i, y_i, \theta_i)$ to locate the QUC *i*. Here, $(x_i, y_i)$ is the coordinate of the center of the QUC in the *xy* plane, and $\theta_i$ is the angle of the QUC orientation with respect to *x* axis. The orientation of a QUC is defined as the vector drawn from the center of the QUC to the center of the octagon in the right in Fig.3(b) which corresponds to $\overrightarrow{OA}$ in Fig.3(a). Next, we consider another QUC *j* with the coordination $(x_j, y_j, \theta_j)$ defined same as above. Thus the relative position of QUC *j* with respect to QUC *i* is determined by three parameters $(d, \alpha, \beta)$, which are defined as

$$d = \sqrt{(x_j - x_i)^2 + (y_j - y_i)^2}$$
$$\alpha = tg^{-1}\left[(y_j - y_i)/(x_j - x_i)\right]$$
$$\beta = \theta_j - \theta_i .$$

If we set the spacing between any two centers of adjacent outer octagons to be 1, then the distance between the centers of two covering QUCs can take three values $d_A = 2\cos(\pi/8) \simeq 1.306$ or $d_B = 1$ or $d_C = \sqrt{2} + 1 \simeq 2.414$. In Fig.5, the distance is $d = d_A$ for the coverings C1-C7, $d = d_B$ for the coverings C8-C9, and $d = d_C$ for the coverings C10-C29. The angles $\alpha$ and $\beta$ are multiples of $\pi/8$. The detailed parameters $(d, \alpha, \beta)$ of each covering type are listed in Table 1.

The QUCs which are covering a single QUC may overlap, and their covering type between any two QUCs must be one of the 29 cases. This will result in restrictions for combined coverings. For example, if a QUC is covered by another one according to type C1, it cannot be covered by other QUCs with the covering types C2, C8, C10, C12, C14, C16 and C17. We call these disallowed coverings the inconsistent coverings about C1. The inconsistent coverings of each covering type are listed at the last column in Table 1.

## 4. Summary

We have studied the geometric properties of the octagonal quasilattice in terms of QUC covering. The two-color QUC is obtained and the covering rule is presented. It is shown that an octagonal quasilattice can be formed by repetitions of a single type QUC. Although there is no rigorous proof for the results, the computer calculations show that a very large patch of octagonal quasilattice obeys the covering rule perfectly. We thus believe the QUC scheme for the octagonal quasilattice is robust. This work extends the Gummelt's decagonal covering theory to the octagonal quasilattice and hopefully the covering theory can be used to describe other quasilattices.

**Acknowledgments**

This work was supported by the National Natural Science Foundation of China under Grant No. 10474021.

**Figure captions**

**Fig.1** The Gähler's clusters for obtaining the octagonal quasilattice. (a) The octagon cluster and the ship cluster. (b) The arrowed octagon cluster. (c) The unarrowed octagon cluster which has the same asymmetry and imposes the same overlapping constraints as the arrowed octagon cluster.

**Fig.2.** Part of a dart-triangle tiling and the clusters used to construct the QUC. (a) Eight overlapped A clusters and the vacancy consisting of eight triangles at the center. (b) The A cluster plus eight octagons outside forms the B cluster. (c) Two vacancies appear between cluster B1 and B2. One vacancy appears between cluster B2 and B3. (d) The B cluster plus eight darts and eight triangles forms the C cluster.

**Fig.3.** (a) The C cluster and its correspondence to Gähler's arrowed octagon (bold lines). (b) The QUC of the octagonal quasilattice.

**Fig.4.** A patch of the octagonal quasilattice generated by covering of the two-color QUC.

**Fig.5.** The 29 covering cases that a C cluster (bold outline) is covered by another one.

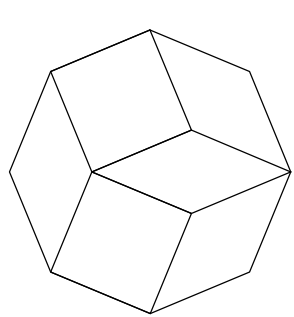 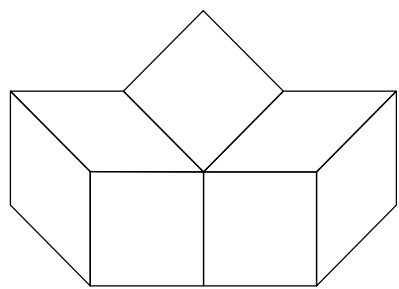 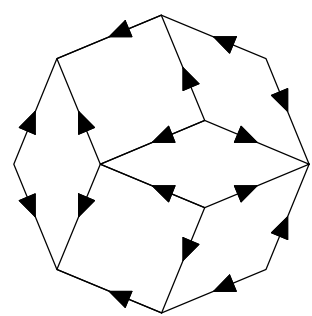 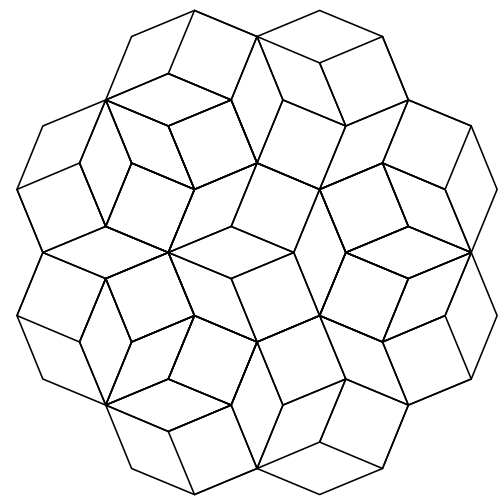

(a)　　　　　　　　　　(b)　　　　　　　(c)

Fig.1

Figure(s)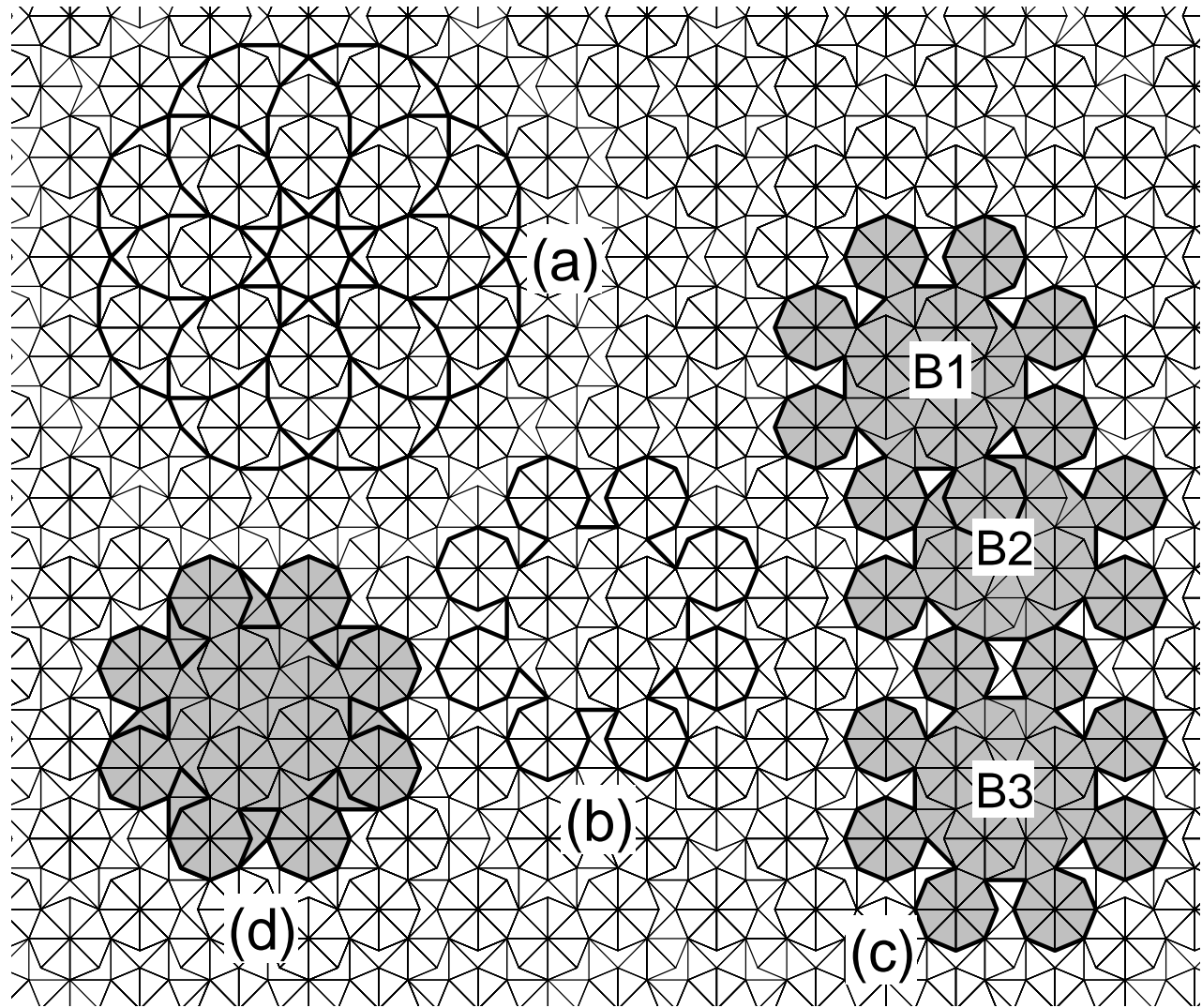

Fig.2

Figure(s)

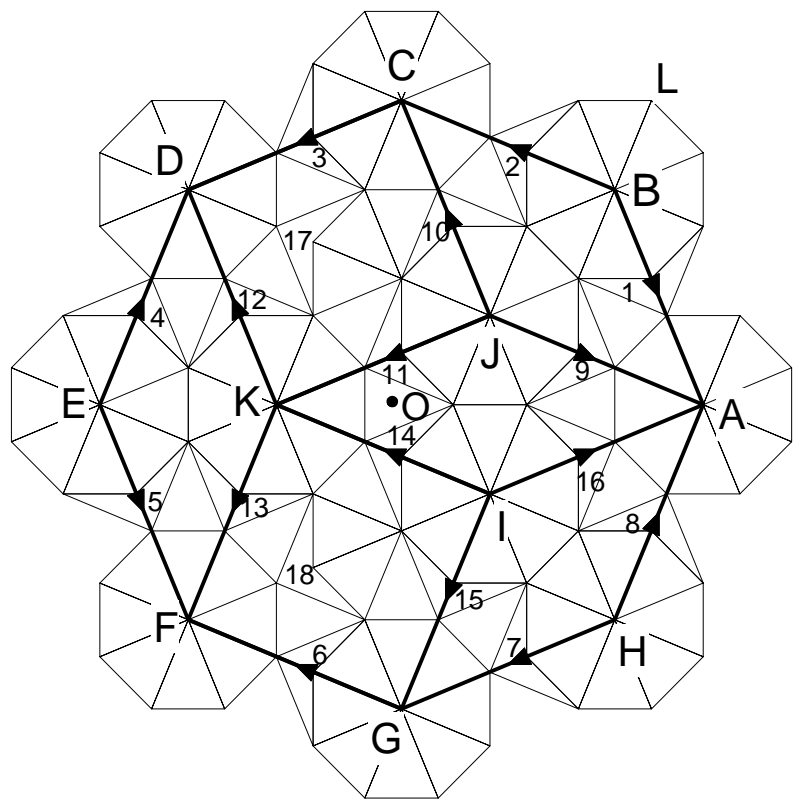
(a)

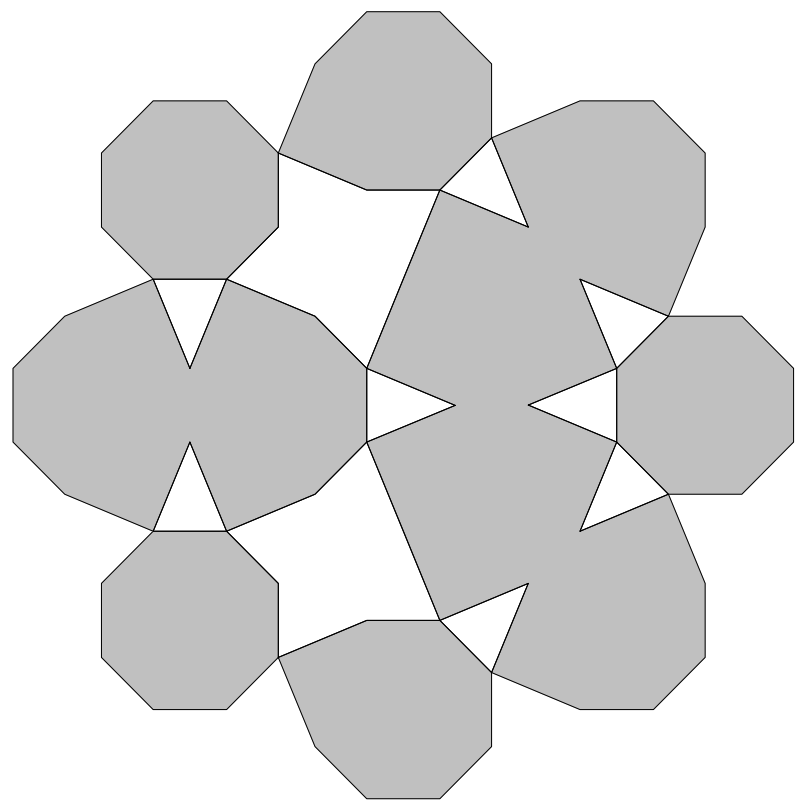
(b)

Fig.3

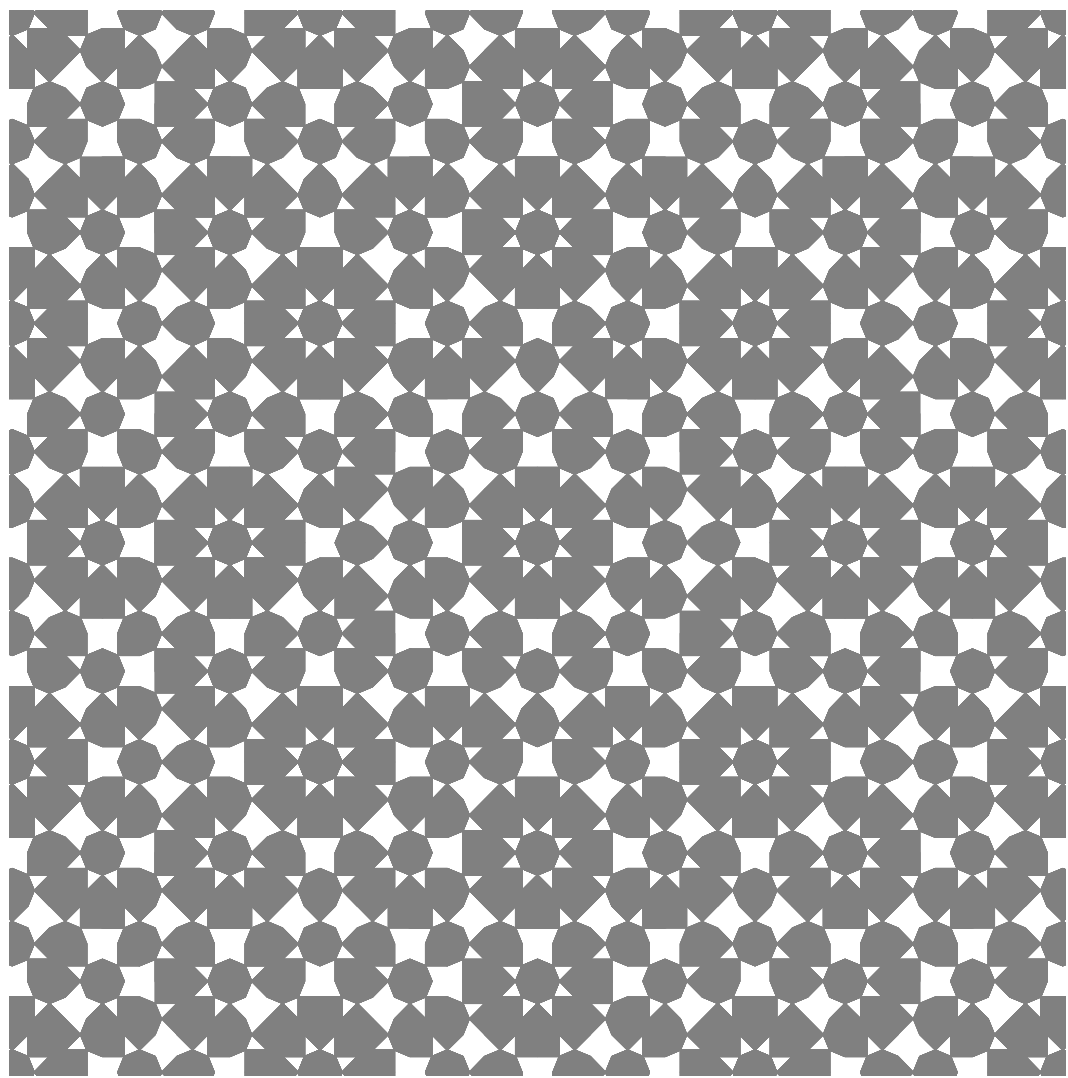

Fig.4

Figure(s)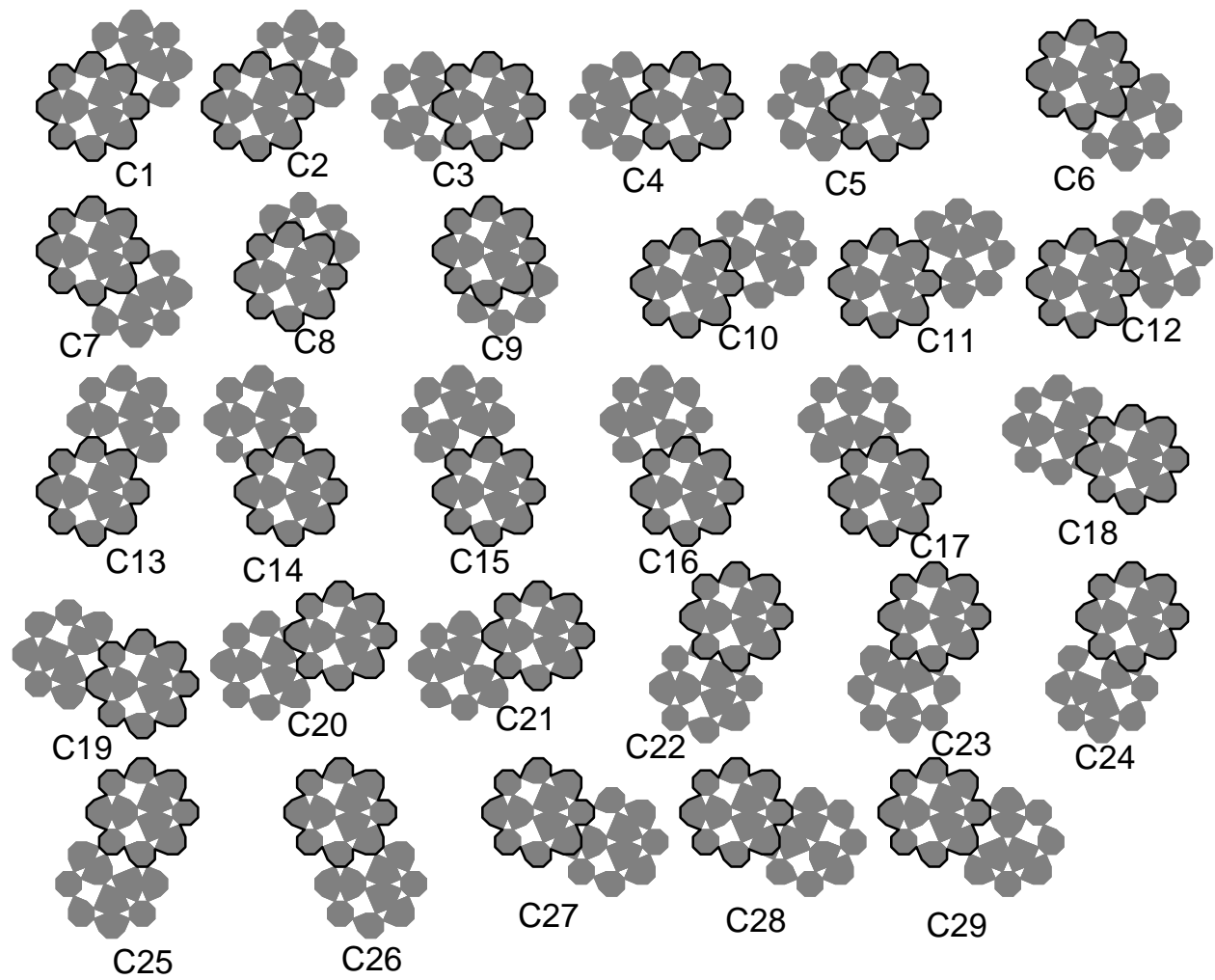

Fig.5

**Table(s)**

Table 1. The parameters of the 29 covering types between two octagonal QUCs.

| Covering type | d | $\alpha$ | $\beta$ | Inconsistent coverings |
|---|---|---|---|---|
| C1 | $d_A$ | $\pi/4$ | $\pi/4$ | C2,C8,C10,C12,C14,C16,C17 |
| C2 | $d_A$ | $\pi/4$ | $3\pi/2$ | C1,C10,C11,C13,C15 |
| C3 | $d_A$ | $\pi$ | $\pi/4$ | C4,C5,C14,C19,C20,C21,C23,C24,C25 |
| C4 | $d_A$ | $\pi$ | $\pi$ | C3,C5,C14,C18,C20,C22 |
| C5 | $d_A$ | $\pi$ | $7\pi/4$ | C3,C4,C15,C16,C17,C18,C19,C21,C22 |
| C6 | $d_A$ | $7\pi/4$ | $\pi/2$ | C7,C25,C26,C27,C29 |
| C7 | $d_A$ | $7\pi/4$ | $7\pi/4$ | C6,C9,C22,C23,C24,C27,C28 |
| C8 | $d_B$ | $3\pi/8$ | $7\pi/4$ | C1,C11,C13,C15 |
| C9 | $d_B$ | $13\pi/8$ | $\pi/4$ | C7,C25,C26,C29 |
| C10 | $d_C$ | $\pi/8$ | 0 | C1,C2,C11,C12,C13,C27,C28,C29 |
| C11 | $d_C$ | $\pi/8$ | $\pi/2$ | C2,C8,C10,C12,C27 |
| C12 | $d_C$ | $\pi/8$ | $5\pi/4$ | C1,C10,C11,C13,C27 |
| C13 | $d_C$ | $3\pi/8$ | 0 | C2,C8,C10,C12,C14,C16,C17 |
| C14 | $d_C$ | $5\pi/8$ | 0 | C1,C3,C4,13,C15,C16,C17,C18,C19 |
| C15 | $d_C$ | $5\pi/8$ | $\pi/4$ | C2,C5,C8,C14,C16,C17 |
| C16 | $d_C$ | $5\pi/8$ | $3\pi/4$ | C1,C5,C13,C14,C15,C17,C18,C19 |
| C17 | $d_C$ | $5\pi/8$ | $3\pi/2$ | C1,C5,C13,C14,C15,C16 |
| C18 | $d_C$ | $7\pi/8$ | 0 | C4,C5,C14,C16,C19,C20,C21 |
| C19 | $d_C$ | $7\pi/8$ | $5\pi/4$ | C3,C5,C14,C16,C18,C20 |
| C20 | $d_C$ | $9\pi/8$ | 0 | C3,C4,C18,C19,C21,C22,C24 |
| C21 | $d_C$ | $9\pi/8$ | $3\pi/4$ | C3,C5,C18,C20,C22,C24 |
| C22 | $d_C$ | $11\pi/8$ | 0 | C4,C5,C7,C20,C21,C23,C24,C25,C26 |
| C23 | $d_C$ | $11\pi/8$ | $\pi/2$ | C3,C7,C22,C24,C25,C26 |
| C24 | $d_C$ | $11\pi/8$ | $5\pi/4$ | C3,C7,C20,C21,C22,C23,C25,C26 |
| C25 | $d_C$ | $11\pi/8$ | $7\pi/4$ | C3,C6,C9,C22,C23,C24 |
| C26 | $d_C$ | $13\pi/8$ | 0 | C6,C9,C22,C23,C24,C27,C28 |
| C27 | $d_C$ | $15\pi/8$ | 0 | C6,C7,C10,C11,C12,C26,C28,C29 |
| C28 | $d_C$ | $15\pi/8$ | $3\pi/4$ | C7,C10,C26,C27,C29 |
| C29 | $d_C$ | $15\pi/8$ | $3\pi/2$ | C6,C9,C10,C27,C28 |